\documentclass[a4paper,12pt]{article}
\usepackage{amssymb}
\usepackage{amsmath}
\usepackage{amscd}
\usepackage{amstext}
\usepackage{ulem}
\usepackage{graphicx} 
\usepackage{rsfs}
\usepackage{latexsym}
\usepackage{amssymb}
\usepackage{graphicx}

\begin{document}
\title{Remarks on the proton charge radius}
\author{Hirohisa Ishikawa 
\footnote{E-mail address: ishikawa@meikai.ac.jp}\\
{\small \it Department of Economics, Meikai University,}\\
 {\small \it Urayasu, Chiba, 279-8550, Japan.}\\
{}
\\
and\\
Keiji Watanabe
\footnote{Present address; 5-36-2 Akazutumi, Setagaya-ku 156-0044}\\
{\small \it Department of Physics, Meisei University,}\\
{\small \it Hino, Tokyo 191-8506, Japan}           }

        \date{}
\maketitle
\abstract{Proton charge radius is calculated from the electromagnetic form factor of proton parameterized by the dispersion relation. 
 The calculated charge radius is a little larger than that obtained by the Lamb shift of the $\mu$ mesic atom. As the result is sensitive to the experimental data of proton 
electric form factor at small momentum transfer, more accurate data are 
required to draw conclusion if the result of the
nucleon form factor is different or not from that obtained Pohl et al. }\\

Recently, very accurate proton charge radius has been reported by Pohl et al. 
 obtained by the measurement of Lamb shift of the $\mu$ mesonic atom \cite{pohl}.  The proton charge radius is smaller than that obtained by the electron proton 
 scattering; they report the proton charge  $r_p=0.84184(67)$ fm, while 
$r_p$ = 0.96276 fm when the proton electric form factor is approximated by
 the 
well-known dipole formula, being very good approximation for the small 
momentum  transfer.
Flowers emphasized the possibility that the discrepancy may lead to the break 
down of QED \cite{flowers}. 

The mean square charge radius obtained by the electron proton scattering is 
given in term of the Fourier transform of the form factor by a simple formula
\begin{equation}
 \langle r^2 \rangle = \lim_{Q^2\to 0}\left[-6
  \frac{d}{dQ^2}F(Q^2)\right].             \label{derive}
\end{equation}
To obtain the  experimental value for the proton charge radius numerical 
differentiation should be performed.  

The experimental data of the electromagnetic form factor are given for small $Q^2$  by Borkowski et al. \cite{borkowski} and  Simon et al. \cite{simon}, 
which are illustrated  Fig.1. To obtain the experimental value 
for the charge radius it necessary to calculate Eq.(\ref{derive}) for the 
experimental data. Sick \cite{sick} 
and De R\'uja \cite{rujula} approximated the experimental data  \cite{simon} by
analytic functions and performed numerical differentiation. They obtained 
the following proton charge radius: 
$\sqrt{\langle r^2 \rangle}$ = 0.895 $\pm$ 0.018 fm. The result is a little 
larger than that of Pohl et al.  Considering the error, De R\'uja conclude the 
that QED is not endangered by the experiment of  Pohl et al. It must be
 noted 
that the numerical differentiation is dependent on how the experimental data 
are approximated.

We have performed analysis of nucleon electromagnetic form factors by using the
 dispersion relation
 \begin{equation}
  F_1(Q^2)=\frac{1}{\pi}\int_{4m_{\pi}^2}^{\infty}ds\frac{{\rm Im}F_1(s)}{s+Q^2},\label{disp}
 \end{equation}
 where $F_1$ is the Dirac form factor, $m_{\pi}$ is the pion mass, $s$ is the 
 squared energy for the nucleon and antinucleon system and $Q^2$ is the 
 squared space-like momentum transfer. 
  $F_1$, given by Eq.(\ref{disp}),  satisfies appropriate properties such as
    low energy hadronic
   properties, the vector boson dominance and QCD constraints \cite{FIW}. 
 We were able to  parameterize the exiting 
electromagnetic form factors of nucleons, for the space-like and the 
time-like momenta as well and very good agreement with the 
experimental data was obtained. It is the purpose of this paper to 
calculate the proton charge radius by our dispersion  relation. 
In our analysis 
we used the data of Borkowsski el al. \cite{borkowski} but omitted 
Simon el al. \cite{simon} as the errors of the latter are random errors 
only. We thought the data 
were not appropriate in performing the chi square test.

Before discussing the numerical results, we give some remarks on the 
form factor. As the well-known dipole formula $G_D=1/(1+Q^2/0.71)^2$, with $Q^2$ expressed 
in terms of GeV${^2}$ represents the experimental data of $G_E^p$ fairly well,  the proton charge radius corresponding to the dipole formula is taken 
as a standard. The  experimental data and the theoretical results as well are
 given as the ratio to $G_D$. When $G_E^p/G_D$ is substituted to 
 Eq.(\ref{derive}), we get the difference $\langle r^2 \rangle|_{G_E^p}-r_D^2$, 
 where $r_D$ is the charge radius corresponding to the dipole formula. If the 
 experimental data $G_E^p/G_D$ becomes larger than 1 near $Q^2\approx0$, 
 $\langle r^2\rangle |_{G_E^p}$ becomes smaller than $r^2_D$. 
 
 We have considered the Sachs electric form factor so far, but we have the 
 Dirac charge form factor $F_1^p$.  It must be noted that the proton charge 
 radius  is to be calculated by using $F_1^p$ in (\ref{derive}). Writing the 
 Charge form factor in terms of the Sachs form factors, we have
 \begin{equation}
  F_1^p(Q^2)=\left(G_E^p+\frac{Q^2}{4m^2}G_M^p\right)
   /\left(1+\frac{Q^2}{4m^2}\right),
\end{equation}
where $m$ is the proton mass. Therefore, 
\begin{equation}
 \langle r^2 \rangle|_{F_1^p}=-6\frac{d}{d Q^2}F_1^p(Q^2)|_{Q^2=0}
   = -6\frac{d}{d Q^2}G_E^p|_{Q^2=0}-\frac{3g^p}{2m^2},
\end{equation}
where $g^p$ is the proton anomalous magnetic moment, $g^p=1.79284939$.
Thus we have
\begin{equation}
  \langle r^2 \rangle|_{F_1^p}=\langle r^2 \rangle|_{G_E^p}-\frac{3g^p}{2m^2}
\end{equation}
in GeV unit.

Let us give our calculations obtained by the dispersion relation \cite{FIW}, 
where the proton data of Simon et al. \cite{simon} were not used as was
 mentioned before. In this paper we take account of Simon et al., and leave out Borkowski 
et al., as the former is more accurate for small $Q^2$. We investigate how 
the proton charge radius change by this modification. 

\begin{figure}[htbp]
\begin{center}
\includegraphics[width=10.0cm]{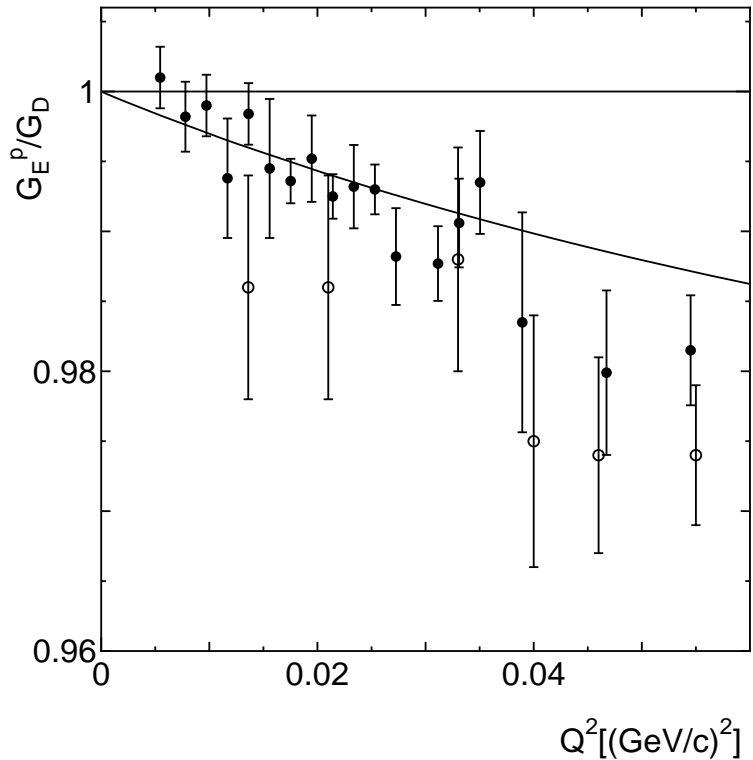}
\label{fig2}
\end{center}
\caption{The  theoretical curve obtained by replacing Borkowski et al. 
\cite{borkowski} by Simon et al. 
\cite{simon}. Closed circles represent Simon et al. \cite{simon} and the open circles Borkowski et al. \cite{borkowski}.}
\end{figure}

We give our numerical values of the proton charge radius calculated by our 
dispersion relation. It must be noted that the data points are the same as in 
Ref. \cite{FIW} except for the proton charge form factor at low $Q^2$. The 
case II of \cite{FIW} is used.

First we give the proton charge radius calculated by using the parameters 
determined in Ref. \cite{FIW}.\\
$\quad  \langle r^2 \rangle |_{G_E^p}$ = 25.751 GeV$^{-1}$ .\\
When the data of Borkowski et al. \cite{borkowski} are omitted and Simon 
et al. \cite{simon} are added we have the following result:\\
 $\quad \langle r^2 \rangle |_{G_E^p}$ = 24.5551 GeV$^{-1}$. 
 
 Our calculated 
 result agree with data of Simon et al. rather than Borkowski et al. The 
 total value of $\chi^2$, addition of space-like and the time-like part, 
  becomes $\chi^2_{tot}$= 505.6 after the replacement of Borkowski 
  et al. by the Simon et al. Our previous value of Ref.\cite{FIW} is  524.5. We illustrate in Fig.1 the calculated result together with experimental data.
  
 The proton charge radius is thus obtained to be
 \begin{equation}
  \sqrt{\langle r^2 \rangle|_{F_1^p}}=0.9401 \,\,{\rm fm},
 \end{equation}
 with the experimental data and the parameters given in \cite{FIW}, and
 \begin{equation}
  \sqrt{\langle r^2 \rangle|_{F_1^p}}=0.9149 \,\,{\rm fm},
 \end{equation}
 where the data of Borkowski et al., are replaced by Simon et al.
 
 The charge radius of proton is a little larger than the result of Pohl et al. 
 However it is premature to conclude that the QED is endangered by their 
 experiment, as the proton charge radius obtained from the electromagnetic 
 form factor is sensitive to the low $Q^2$ part of the form factor. More 
 accurate data near $Q^2\approx 0$ are required.

\end{document}